# PULSED ELECTRON SOURCE CHARACTERIZATION WITH THE MODIFIED THREE GRADIENT METHOD

S. Marghitu, C. Oproiu, NILPRP, Acc. Lab., Bucharest–Magurele, R–76900, Romania
D. C. Dinca, MSU–NSCL, East Lansing, MI 48824–1321, USA
O. Marghitu, NILPRP – ISS, Bucharest–Magurele, R–76900, Romania

Abstract

Results from the Modified Three Gradient Method (MTGM), applied to a pulsed high intensity electron source, are presented. The MTGM makes possible the non–destructive determination of beam emittance in the space charge presence [1]. We apply the MTGM to an experimental system equipped with a Pierce convergent diode, working in pulse mode, and having a directly heated cathode as electron source. This choice was mainly motivated by the availability of an analytical characterization of this source type [2], as well as the extended use of the Pierce type sources in linear accelerators. The experimental data are processed with a numerical matching program, based on the K–V equation for an axially symmetric configuration [3], to determine the emittance and object cross–over position and diameter. The variation of these parameters is further investigated with respect to both electrical and constructive characteristics of the source: cathode heating current, extraction voltage, and cathode–anode distance.

## 1 INTRODUCTION

Non–destructive measurements of emittance and object cross–over radius and position, for a pulsed high intensity electron source, are presented. Our goal is to check the possibility of using MTGM as a reliable routine tool in beam diagnosis.

The MTGM, [1], is based on three gradient type measurements of the beam cross–section and on the subsequent use of a computer code incorporating the K–V equation. The experimental installation built up to set and check the method is shown as Figure 1.

## 2 THEORETICAL BACKGROUND

The envelope of an axially symmetric beam, propagating in an electric field free region, in paraxial approximation, follows the equation (e.g. [3]):

$$\frac{d^2 R}{dz^2} + \frac{\eta}{8} \frac{B^2}{V} R = \frac{1}{4\pi\varepsilon_0} \frac{I}{V^{3/2}} \frac{1}{R} + \frac{\varepsilon^2}{R^3} \quad (1)$$

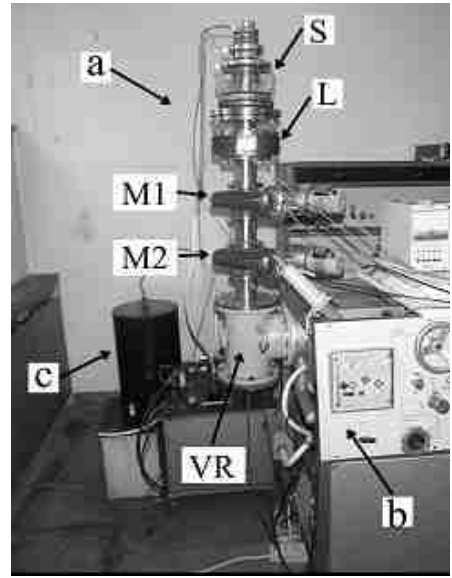

Figure 1: Experimental set–up
a – beam system; b – vacuum installation; c – pulse high voltage transformer. The beam system consists of: S – the electron source; L – thin, axially symmetric, magnetic lens; M1, M2 – beam profile monitors (BPM); VR – vacuum room, with a specially designed Faraday cup inside where: $\eta$ = charge–to–mass ratio for the electron, $\varepsilon_0$ = dielectric constant, I = beam current, V = beam acceleration potential, $\varepsilon$ = beam emittance (which according to Liouville's theorem remains constant), R = beam envelope, and B = axial magnetic field.

To solve equation (1) is necessary to know the parameters I, V, B, and $\varepsilon$, as well as some initial conditions. It turns out that only $\varepsilon$ requires a special effort to be determined; I and V can be directly measured, whereas B=B(z) depends on the geometry of the lens and on its polarization, and can be calculated with dedicated software.

The initial conditions are also unknown. A good choice for us is the distance of the object cross–over from the center of the focusing lens L, $z_0$, and its radius, $R_0$. Consequently, to get the evolution of the beam, one has to find ($\varepsilon$, $z_0$, $R_0$).

# 3 MEASUREMENTS

As already mentioned, V and I are measured directly, by using a two channel digital oscilloscope. An example oscillogram is given here as Figure 2. The 'shorter' pulse is the current, I, at M1 exit plane, measured on a 1Ω resistor, while the 'longer' one is the high−voltage, V. For the example shown V=31.7kV, I=0.43A. The corresponding cathode heating current and lens polarization voltage are $I_{fil}$=8.4A and $U_L$=4.4V.

Before proceeding to beam cross−section measurements, the volt−ampere characteristics of the source was obtained, for two different geometries, IG (initial geometry) and MG (modified geometry). For IG the distance between the anode tip and the emissive filament is $d_{ac1}$ =19mm, whereas for MG $d_{ac2}$ =22mm. The oscillogram in figure 2 corresponds to geometry IG. The function $I=I(V,I_{fil})$ is tabulated below, in Tables 1 and 2, for respectively IG and MG.

Table 1− Current beam I [A] at anode exit for IG.

| V[kV] $I_{fil}$ [A] | 9.3 | 18.6 | 27.9 | 37.2 | 46.5 | 55.8 |
|---|---|---|---|---|---|---|
| 8.1 | 0.13 | 0.228 | 0.264 | 0.284 | 0.3 | 0.32 |
| 8.6 | 0.15 | 0.38 | 0.5 | 0.536 | 0.61 | 0.63 |
| 9.1 | 0.2 | 0.48 | 0.712 | 0.96 | 1.1 | 1.18 |
| 9.5 | – | 0.484 | 0.8 | 1.16 | 1.5 | 1.7 |

Table 2− Current beam I [A] at anode exit for MG

| V[kV] $I_{fil}$ [A] | 10.4 | 15.6 | 20.8 | 26.0 | 31.6 | 36.4 |
|---|---|---|---|---|---|---|
| 8.4 | 0.102 | 0.168 | 0.22 | 0.26 | 0.32 | 0.33 |
| 8.6 | 0.11 | 0.19 | 0.24 | 0.31 | 0.37 | 0.44 |
| 8.8 | 0.114 | 0.198 | 0.28 | 0.35 | 0.42 | 0.5 |
| 9.1 | 0.116 | 0.2 | 0.29 | 0.39 | 0.46 | 0.6 |

Determination of the beam diameter is the most sensible part of the measurements, [4]. Each BMP consists of wire scanner that crosses the beam at constant velocity, $v_M$. The diameter results by multiplying the velocity with the scanning time, $\tau_M$, as read with a second oscilloscope. The measurements have to be conducted with great care, because of various potential error sources; in particular, for the low energy range emphasized here, the backscattered electrons can seriously alter the data. Upper part of figure 3 shows the pulses obtained when a diaphragm in front of the Faraday cup is placed too close to M1; lower part of the figure shows the effect of removing the diaphragm. These data correspond again to geometry IG.

For each case studied the beam radii, R1 and R2, are measured as function of the lens polarization voltage, $U_L$. The experimental values are then fitted with polynomials (3 to 5 degree), and the coefficients fed to the computer code MTGMprog, developed to assist MTGM. The program is based on a Monte Carlo algorithm that searches the (ε, $z_0$, $R_0$) parameter space, until the best fit to the data, within a given error, is found. A typical result is given in the next section, as figure 7.

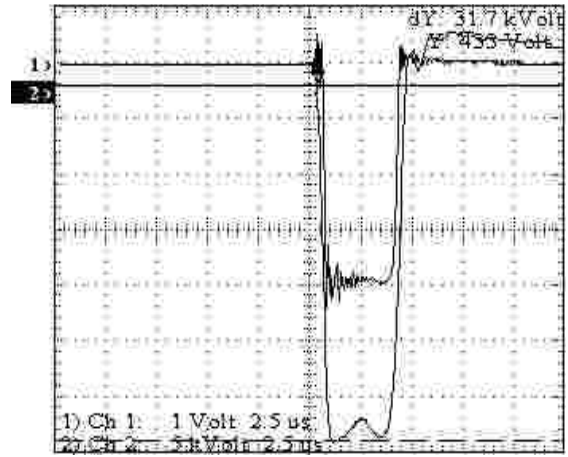

Figure 2: Oscillogram of V ('longer' pulse) and I ('shorter' pulse) at M1 exit plane; $I_{fil}$=8.4A, $U_L$ =4.4V

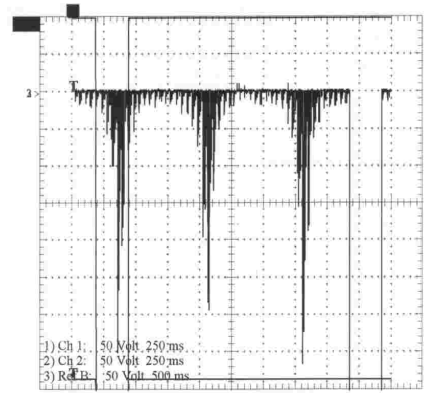

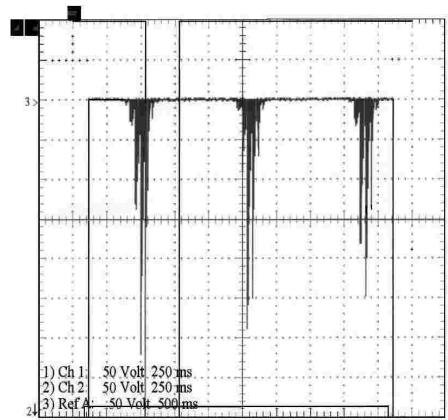

Figure 3: Effect of the back−scattered electrons on the beam cross−section determination

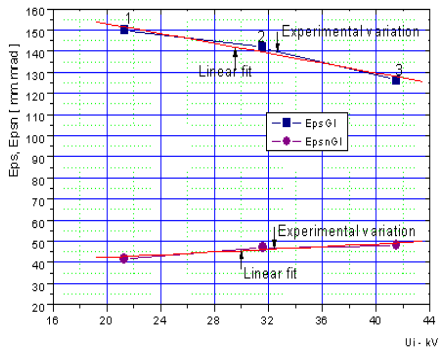

Figure 4: Emittance variation for IG; $I_{fil}$=ct=8.4A; $I_1$, $I_2$, $I_3$ are: 0.32A, 0.41A, 0.46A

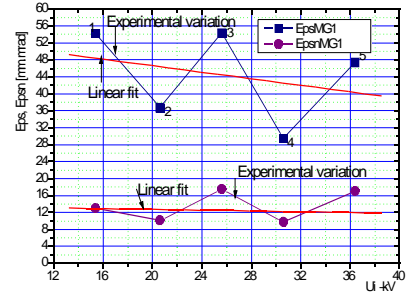

Figure 6: Emittance variation for MG; $I_{fil}$=ct=8.4A; $I_1$, $I_2$, $I_3$, $I_4$, $I_5$ are: 0.19A, 0.26A, 0.34A, 0.42A, 0.52A

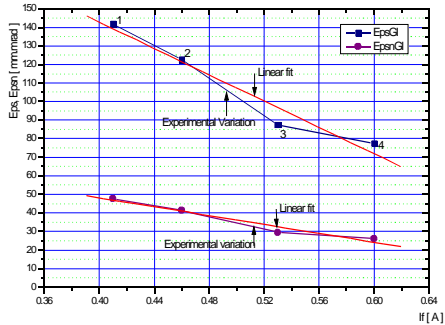

Figure 5: Emittance variation for IG; V=ct=31.6kV; $I_{fil1}$, $I_{fil2}$, $I_{fil3}$, $I_{fil4}$ are: 8.4A, 8.5A, 8.6A, 8.7A

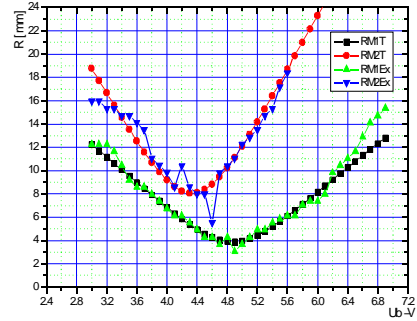

Figure 7: Beam radii $R_1$, $R_2$ dependence on $U_L$; experimental measurement vs. numerical fit

## 4 RESULTS

The dependence of the emittance, $\varepsilon$, on V and $I_{fil}$, for the geometry IG, is shown in figures 4 and 5 (note that in figure 5 the beam current, I, is used for the abscissa). In particular, figure 5 corresponds to the usual case in linear electron accelerators, with V fixed by design. One can see that the variation of the normalized emittance, $\varepsilon_n$, is rather small, in agreement with the theory. Another observation refers to the relatively linear variation of $\varepsilon$ with respect to both parameters, which makes possible the use of a linear fit, once a few experimental poins were determined.

Figure 6 presents the dependence of $\varepsilon$ on V for the geometry MG. The most pregnant feature, compared to figure 4, is the large variation (~300%) for a relatively small (~16%) change in the anode−cathode distance. This variation is in the expected sense: the larger is the distance $d_{ac}$, the more uniform is the electric field in between and smaller the $\varepsilon$. The variation of $\varepsilon_n$ is again small, and the trend linear, although this time there is a significant scatter of the points. This is probably related to the computing code.

Because of lack of space we cannot show here graphs with the variation of the object cross−over position and radius, $z_0$ and $R_0$. However, consistency checks between the measured and calculated data were performed. An example is shown in figure 7, that corresponds to point 2 in figure 4. For this case $R_0$=1.71mm, $z_0$=73.8mm. The good match between the measured and calculated values is evident.

To conclude, we consider the results presented here as very promising. Further work is needed to improve the computing code, as well as for accumulating a better case statistics.

*Acknowledgment*: Work supported by the Romanian Ministry of Education and Research, grant 3216C/2000.